# Distributed Stochastic Model Predictive Control for Human-Leading Heavy-Duty Truck Platoon

Mehmet Fatih Ozkan, *Graduate Student Member, IEEE*, and Yao Ma, *Member, IEEE*

*Abstract*—Human-leading truck platooning systems have been proposed to leverage the benefits of both human supervision and vehicle autonomy. Equipped with human guidance and autonomous technology, human-leading truck platooning systems are more versatile to handle uncertain traffic conditions than fully automated platooning systems. This paper presents a novel distributed stochastic model predictive control (DSMPC) design for a human-leading heavy-duty truck platoon. The proposed DSMPC design integrates the stochastic driver behavior model of the human-driven leader truck with a distributed formation control design for the following automated trucks in the platoon. The driver behavior of the human-driven leader truck is learned by a stochastic inverse reinforcement learning (SIRL) approach. The proposed stochastic driver behavior model aims to learn a distribution of cost function, which represents the richness and uniqueness of human driver behaviors, with a given set of driver-specific demonstrations. The distributed formation control includes a serial DSMPC with guaranteed recursive feasibility, closed-loop chance constraint satisfaction, and string stability. Simulation studies are conducted to investigate the efficacy of the proposed design under several realistic traffic scenarios. Compared to the baseline platoon control strategy (deterministic distributed model predictive control), the proposed DSMPC achieves superior controller performance in constraint violations and spacing errors.

*Index Terms*—Truck platooning, intelligent transportation systems, driver behavior, inverse reinforcement learning, stochastic model predictive control.

## I. INTRODUCTION

TRUCK platooning has gained considerable attention in recent decades with the growing technology readiness in intelligent transportation systems (ITS). The main objectives of the truck platooning systems are to improve road utilization, safety, fuel economy, and carbon emissions in traffic [1-3]. With the recent developments of the ITS, controlling an automated truck platoon in a form is feasible with different control strategies such as distributed and centralized formation control approaches [4-5].

Despite the growing vehicle connectivity and autonomy in the truck platooning systems, the highly uncertain and dynamic traffic conditions still prove to be challenging for the fully automated truck platoons to operate in all circumstances safely. As an alternative, a supervising human driver at the leader truck has been proposed to enhance the platoon's capability to handle unforeseen situations [6-7]. This human-leading platooning system has the advantages of human expertise that is assisted with vehicle connectivity and autonomy, compared to the existing fully automated truck platooning systems [8]. With the human-leading truck platooning system, the human-driven leader truck can strategically deal with traffic conditions that are difficult for existing fully automated trucks, such as construction zones, traffic control situations, erratic drivers, weather changes, etc., while guiding the fully automated following trucks on the road [9-10]. This proposed implementation effectively synthesizes vehicle autonomy with human intelligence, which substantially expands and accelerates the application and adoption of automated trucks in realistic traffic situations.

In the human-leading truck platooning systems, the following automated trucks should accurately predict the intent of the human-driven leader truck to control the platoon in a form safely. However, the stochastic nature of a human's decision-making mechanism challenges understanding the intentions of the human-driven leader truck by the following automated trucks in traffic. Therefore, it is essential to develop an accurate driver behavior model that captures the stochasticity of the driver behavior in real-world driving scenarios. Existing studies have widely used inverse reinforcement learning (IRL) based driver behavior models to explain human's internal decision-making strategies when operating vehicles [11-15]. The IRL approach assumes that human drivers are rational decision-makers whose aims are selecting driving maneuvers to minimize cost functions [15]. The cost function encodes the unique driving preferences of the human driver, such as desired speed and car-following gap distance. The IRL intends to recover this cost function from driver-specific demonstrations and recursively solve the optimal control problem with respect to the learned cost function to compute personalized driver-specific trajectories. However, most traditional IRL-based driver behavior models are deterministic and aim to obtain a single cost function from a particular collection of driving demonstrations. In practice, this learned single cost function is unable to sufficiently describe the stochasticity of the human driver behavior. Over the past decade, some IRL-based learning models have been proposed to learn multiple cost functions in different applications [16-19]. A recent probabilistic IRL-based driver

Mehmet Fatih Ozkan and Yao Ma are with the Department of Mechanical Engineering, Texas Tech University, Lubbock, TX, 79409 (e-mail: mehmet.ozkan@ttu.edu; yao.ma@ttu.edu).

behavior model is proposed to learn multiple cost functions from a given collection of demonstrations where each cost function is learned for a particular driver [19]. This suggested driver behavior model considers the uniqueness of the driver behavior, but the time-varying stochastic nature of human driver behavior is not explicitly addressed.

In the vehicular platooning systems, the distributed model predictive control (DMPC) approach has been widely used to control the platoon by maintaining constant inter-platoon gap distances [20] or constant time headway between the trucks [21] with the string stability property [22-23]. The DMPC approach enables each vehicle in the platoon to be effectively controlled online while addressing the input and state constraints and predicting the evolution of the system with time. However, most existing DMPC strategies do not consider the stochastic disturbances affecting the system and assume accurate predictions of the predecessors' state information in the platoon. In the human-leading truck platoon system, the stochastic driving behavior of the human-driven leader truck may lead the existing DMPC approaches to take wrong decisions that incur safety constraint violations, recursive infeasibility, and string instability in the platoon. Therefore, some studies suggest that the vehicle control problem can be solved with the stochastic model predictive control (SMPC) approach to effectively deal with the stochastic driving actions of the preceding vehicle in a variety of traffic situations [24-25]. Moreover, distributed SMPC (DSMPC) approaches have been proposed as an alternative to the DMPC approach for improving the controller performance to efficiently deal with stochastic disturbances in the distributed systems [26-28]. In [26], the authors propose a DSMPC approach with theoretical guaranteed recursive feasibility and closed-loop chance constraint satisfaction for a group of linear systems with stochastic disturbances and chance constraints. Results show that the proposed method can effectively deal with stochastic disturbances while achieving recursive feasibility and closed-loop chance constraint satisfaction. Compared to the developed SMPC approaches, few DSMPC methods for the vehicle control problems have been presented in the literature. In [28], the authors propose a DSMPC approach for a vehicular platoon, and the results demonstrate that the proposed model can obtain string-stable and safe longitudinal control for the platoon by the proper model settings. No theoretical analysis of the recursive feasibility, chance constraint satisfaction, and string stability is provided, though.

Motivated by the aforementioned discussions, this study aims to establish a distributed stochastic model predictive control design integrated with the stochastic driver behavior prediction model to control a human-leading heavy-duty truck platoon in a form during the trip. The distinct contributions of this study include:
1. A stochastic driver behavior learning model for the human-driven leader truck is proposed to learn and predict the driver's stochastic driving strategies in a variety of longitudinal driving scenarios.
2. A distributed stochastic control design for the following automated trucks is implemented with sufficient conditions to satisfy the recursive feasibility, closed-loop chance constraints, and string stability.

To the best of the authors' knowledge, this is the first study explicitly addressing the stochastic driver behavior learning and distributed stochastic formation control for human-leading truck platooning systems.

The rest of this paper is organized as follows. Section II presents the stochastic driver behavior model. Section III formulates the distributed stochastic model predictive control design with the recursive feasibility, chance constraint satisfaction, and string stability analysis. Section IV provides the simulation results along with performance analysis. Section V concludes the paper.

## II. STOCHASTIC DRIVER BEHAVIOR LEARNING

In this study, a human-leading heavy-duty truck platoon with a predecessor-following (PF) communication topology is considered. The platoon consists of a human-driven leader truck $(i=0)$ and three homogenous following automated trucks $(i=1,2,3)$, as shown in Fig. 1.

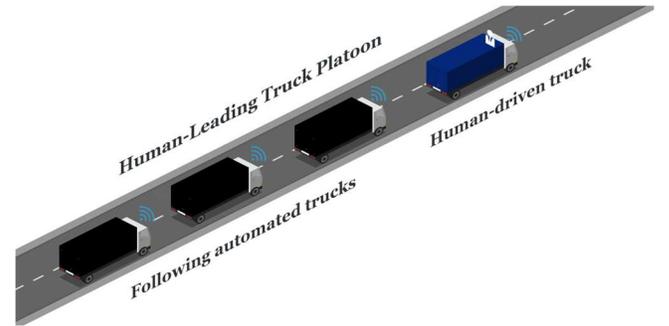

Fig. 1. Schematic of the human-leading heavy-duty truck platoon in traffic.
*Source*: This figure was generated at https://icograms.com.

In this section, the stochastic driver behavior learning model of the human-driven leader truck will be presented. The proposed stochastic driver behavior learning model uses an inverse reinforcement learning approach and aims to learn a cost function distribution that encodes the uniqueness and richness of the truck driver with given demonstrated driving data. The learned cost function distribution of the driver will be used to predict the human-driven leader truck's longitudinal acceleration distribution by the immediate following automated truck during the trip. In the remainder of this section, the implementation details of the proposed stochastic driver behavior learning model will be presented.

### A. Vehicle Trajectory Modeling

In the stochastic driver behavior learning model, the vehicle trajectory is formulated as the longitudinal position of the heavy-duty truck. In general, the space of such vehicle trajectories has infinitely many dimensions. To overcome this problem, a one-dimensional quintic polynomial is used as a finite-dimensional representation of the longitudinal position of the heavy-duty truck in this study. For vehicle trajectory modeling, the quintic polynomials have been widely used in existing studies [11-14] for the benefits of smooth motion, easy calibration, and light computation.

The longitudinal position of the heavy-duty truck is formulated in a time interval $[t_j, t_j + T_H - \Delta t_s]$, $j = 0, 1, \ldots N-1$, for a trajectory that consists of $N$ segments where $T_H$ and $\Delta t_s$ define the length of each trajectory segment and the sample time, respectively. The longitudinal position for each trajectory segment $j$ is defined as

$$r_j(t) = \alpha_0 t^5 + \alpha_1 t^4 + \alpha_2 t^3 + \alpha_3 t^2 + \alpha_4 t + \alpha_5 \quad (1)$$

where $\alpha_{0-5}$ are the polynomial coefficients for each demonstrated trajectory segment and $t \in [0, \Delta t_s, 2\Delta t_s, \ldots, T_H - \Delta t_s]$. The longitudinal velocity and acceleration can be defined as $\dot{r}_j(t)$ and $\ddot{r}_j(t)$, respectively. By utilizing $r_j(t)$, $\dot{r}_j(t)$ and $\ddot{r}_j(t)$ for each trajectory segment, $\alpha_{3-5}$ can be set with the observed longitudinal acceleration, velocity, and position of the heavy-duty truck at $t = 0$, respectively and $\alpha_{0-2}$ can be found through the optimization which will be later mentioned in Algorithm 1.

### B. Stochastic Inverse Reinforcement Learning (SIRL)

In this study, the stochastic inverse reinforcement learning framework (SIRL) is proposed to learn the human driver behavior model from the driver-specific demonstrated driving data. By using a collection of trajectory demonstrations $D$, the objective is to learn a driver's cost function distribution where each cost function in the distribution describes the driver's behavior for each observed trajectory segment. Each cost function is specified as weighted features

$$J_j = \theta_j^T \mathbf{f}_j(r_j) \quad (2)$$

where subscript $j$ represents the $j$th trajectory segment, $J$ is the cost function, $\theta$ is the weight vector, and $\mathbf{f}(r) = (f_1, f_2, \cdots, f_n)^T$ is the feature vector where $n$ defines the number of features. The objective is to find the optimal $\theta^*$ that maximizes the posterior likelihood of the demonstrations for each trajectory segment

$$\theta_j^* = \arg\max_{\theta_j} P_r(D_j | \theta_j) = \arg\max_{\theta_j} \prod_{k=1}^{L} P_r(r_j | \theta_j) \quad (3)$$

where $L$ represents the number of the planning subsegments for each demonstrated trajectory segment and each trajectory subsegment has the same length $T_P$, $P_r(r|\theta)$ represents the probability distribution over the trajectory segment, which is proportional to the negative exponential costs obtained along the trajectory segment based on the Maximum Entropy principle [29] as shown below

$$P_r(r_j | \theta_j) = \exp(-\theta_j^T \mathbf{f}(r_j)) \quad (4)$$

Although the feature weight $\theta$ is usually not derivable analytically, the gradient of the optimization problem with respect to $\theta$ can be derived. The gradient can be calculated by subtracting the observed and predicted feature values [30] as

$$\nabla \mathbf{f}_j = \tilde{\mathbf{f}}_j - \mathbf{f}_j^e \quad (5)$$

where $\tilde{\mathbf{f}}$ is the average observed feature values of the demonstrated trajectory segment $\tilde{r}_j$ as shown below

$$\tilde{\mathbf{f}}_j = \frac{1}{L} \sum_{k=1}^{L} \mathbf{f}(\tilde{r}_{j,k}) \quad (6)$$

The expected feature values can be computed as feature values of the most likely trajectory segment by maximizing $P_r(r_j | \theta_j)$ [11]

$$\mathbf{f}_j^e \approx \mathbf{f}_j\left(\arg\max P_r(r_j | \theta_j)\right) \quad (7)$$

The feature weight vector for each trajectory segment can be updated based on the normalized gradient descent method (NGD) [31] with the learning rate $\gamma$ as in (8)

$$\theta_j \leftarrow \theta_j - \gamma \frac{\nabla \mathbf{f}_j}{\|\nabla \mathbf{f}_j\|} \quad (8)$$

By utilizing the outlined steps above, a collection of $N$ distinct cost functions for a given set of trajectory demonstrations $D$ will be obtained. The learned set of cost functions will be used to generate a distribution in the following stage. To accomplish this, it is crucial to select an efficient distribution model to fit the learned set of cost functions. Copulas are a useful way for constructing a multivariate distribution by breaking down the joint cumulative distribution into marginal cumulative distributions and the copula function. Moreover, copula functions allow modeling dependency between variables that do not follow the same distributions. Considering the learned feature weights do not follow the same distributions and the dependency between the learned feature weights, the learned set of feature weight vectors $\theta = [\theta_1, \theta_2, \theta_3, \ldots, \theta_N]$ is then fitted using t-copula in a multivariate distribution $G_\theta$ [32].

In the t-copula fitting, initially, the kernel density estimation (KDE) [33] method is used to transform the learned feature weight vectors into the copula scale [0,1]. The scaled learned feature weight vectors are then fitted to the t-copula by using the maximum likelihood method, as expressed in [32]. After the fitting process, the samples can be generated from the t-copula, and then the generated samples can be transformed back to the original scale by using the inverse KDE method.

### C. Feature Construction

Since we focus on driving behaviors in the longitudinal direction in this study, the following features are defined to represent the critical factors of the longitudinal driving preferences:

**Acceleration:** The integration of acceleration across each planning horizon is carried out to determine the driver's desired acceleration and deceleration actions within the trajectory subsegment.

$$f_a(t) = \int_t^{t+T_P} \|\ddot{r}(t)\|^2 dt \quad (9)$$

**Desired Speed:** The driver's preferred speed during the trip is determined by integrating the deviation from the cruising speed. The desired speed $v_d$ is defined as the observed maximum speed of the preceding traffic in the trajectory subsegment.

$$f_{ds}(t) = \int_t^{t+T_P} \|v_d - \dot{r}(t)\|^2 dt \quad (10)$$

**Relative Speed:** The driver's desire for following the preceding vehicle speed $v_{PV}$ is used to capture through relative speed integration.

$$f_{rs}(t) = \int_t^{t+T_P} \|v_{PV}(t) - \dot{r}(t)\|^2 dt \quad (11)$$

**Steady Car-following Gap Distance:** The driver's preferred car-following gap distance $d(t)$ is determined by integrating the gap distance deviation from the desired value $d_c$ where $d_s$ is the minimum car-following gap distance and $\varpi$ is the time headway. $\varpi$ is defined as the observed average time headway within the trajectory subsegment.

$$d_c = \dot{r}(t)\varpi + d_s \quad (12)$$

$$f_{cd}(t) = \int_t^{t+T_P} \|d(t) - d_c\|^2 dt \quad (13)$$

**Safe Interaction Gap Distance:** Under the congested traffic condition, the integration of the gap distance variation from the minimum safety car-following gap distance is employed to capture the driver's desired safe interaction gap distance.

$$f_{sd}(t) = \int_t^{t+T_P} \|d(t) - d_s\|^2 dt \quad (14)$$

**Free Motion Gap Distance:** When the driver controls the vehicle in free motion rather than interacting with the preceding vehicle, the integration of the negative exponential growth of the gap distance is utilized to capture the driver's desired gap distance to the preceding vehicle.

$$f_{fd}(t) = \int_t^{t+T_P} e^{-d(t)} dt \quad (15)$$

### D. Feature Selection

In this study, the trajectory segments are clustered depending on the observed driving conditions, and a distinct set of features are applied to each cluster. With this, three different driving conditions are addressed in the longitudinal direction, and the relevant features are applied for each driving condition. To characterize the driving behavior phases for each trajectory segment, the average time headway (THW) and inverse time-to-collision (TTCi) are employed as main indicators. These three distinct driving conditions are outlined and detailed further below.

**Steady car-following:** The steady car-following phase occurs when the average THW < 6 s [34] and average TTCi < 0.05 s$^{-1}$ [35] for each trajectory segment. The features $f_a$, $f_{ds}$, $f_{rs}$ and $f_{cd}$ are used to capture the driver's steady car-following behavior within the trajectory segment.

**Free motion:** During the free motion driving phase, the driver operates the vehicle without engaging with the preceding vehicle on the road. The free motion driving phase within the trajectory subsegment is described using the criteria listed below

1. Average THW $\geq$ 6 s and average TTCi $\leq 0$ s$^{-1}$
2. The average gap distance to the preceding vehicle $\geq 50$ m
3. The average driver's speed $\geq 8$ m/s

The first condition indicates that the driver is not approaching the preceding vehicle, and the driver desires larger time headway to the preceding vehicle during the trip. However, these may occur in congested traffic when the driver closely follows the preceding vehicle at a reduced speed with frequent stop-and-go actions, or in free motion when the driver is not fully impacted by the preceding vehicle. By using the trajectory segments from the driving data, which will be introduced in section IV-A, the trajectory segments that match the first requirement are selected. A K-means clustering algorithm [36] is then utilized to determine whether the driver is in congested traffic or free motion for each trajectory segment by using the selected trajectory segments that match the first condition. The trajectory segments are clustered in the average speed and gap distance space by the K-means algorithm. Fig. 2 shows the clustering results in the average speed and gap distance space. Notably, the trajectories are grouped into two clusters represented by free motion and non-free motion trajectory segments, respectively. By examining the K-means clustering results, the second and third criteria are obtained accordingly. The features $f_a$, $f_{ds}$ and $f_{fd}$ are used to capture the driver's free motion driving strategy for each trajectory segment.

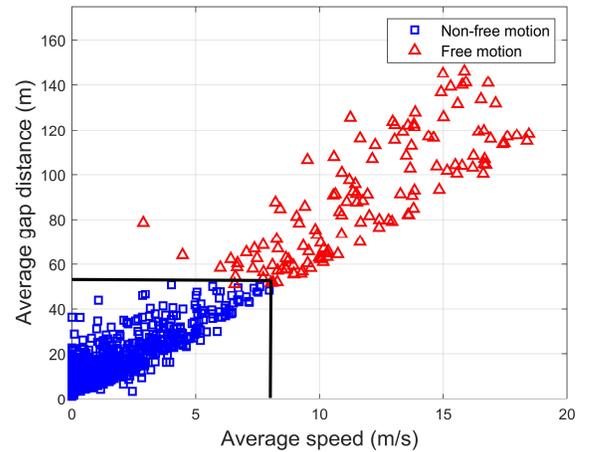

Fig. 2. The clustering results for the free motion driving.

**Unsteady car-following:** The unsteady car-following phase occurs when the driver is neither in steady car-following nor in free motion driving conditions. The features $f_a$, $f_{ds}$, $f_{rs}$ and $f_{sd}$ are used to capture the driver's unsteady car-following behavior within the trajectory segment.

*E. Algorithm Implementation*

Given a set of trajectory demonstrations $D$ consists of $N$ observed trajectory segments $(\tilde{r}_1, \tilde{r}_2, \ldots \tilde{r}_N)$, each trajectory segment is divided into $L$ planning subsegments $(\tilde{r}_{1,1}, \tilde{r}_{1,2}, \ldots, \tilde{r}_{1,L}, \tilde{r}_{2,1}, \tilde{r}_{2,2}, \ldots, \tilde{r}_{2,L}, \ldots, \tilde{r}_{N,1}, \tilde{r}_{N,2}, \ldots, \tilde{r}_{N,L})$. As illustrated in Algorithm 1, the detailed steps are used for the driver behavior learning process.

---

**Algorithm 1: Stochastic driver behavior learning algorithm**

**Input:** $(\tilde{r}_{1,1}, \tilde{r}_{1,2}, \ldots, \tilde{r}_{1,L}, \tilde{r}_{2,1}, \tilde{r}_{2,2}, \ldots, \tilde{r}_{2,L}, \ldots, \tilde{r}_{N,1}, \tilde{r}_{N,2}, \ldots, \tilde{r}_{N,L})$

**Output:** $G_\theta = [G_{\theta_1}, G_{\theta_2}, G_{\theta_3}]$,
$(r_{1,1}^*, r_{1,2}^*, \ldots, r_{1,L}^*, r_{2,1}^*, r_{2,2}^*, \ldots, r_{2,L}^*, \ldots, r_{N,1}^*, r_{N,2}^*, \ldots, r_{N,L}^*)$

1: Cluster the trajectory segments with respect to the driving conditions
2: **for** each cluster **do**
3:   Initialize weight set $\theta \leftarrow [\ ]$
4:   **for** all trajectory segments **do**
5:     $\theta_j \leftarrow$ all-ones vector
6:     $\tilde{\mathbf{f}}_j = \frac{1}{L}\sum_{k=1}^{L} \mathbf{f}_j(\tilde{r}_{j,k})$
7:     **while** $\theta_j$ not converged **do**
8:       **for** all $r_{j,k}^* \in (r_{j,1}^*, r_{j,2}^*, \ldots, r_{j,L}^*)$ **do**
9:         $(\alpha_5, \alpha_4, \alpha_3) \leftarrow$ (position, velocity, acceleration)
10:        at the initial state of the $\tilde{r}_{j,k}$
11:        Optimize $(\alpha_2, \alpha_1, \alpha_0)$ with respect to $\theta_j^T \mathbf{f}_j$
12:      **end for**
13:      $\mathbf{f}_j^e = \frac{1}{L}\sum_{k=1}^{L} \mathbf{f}_j(r_{j,k}^*)$
14:      $\Delta \mathbf{f}_j = \tilde{\mathbf{f}}_j - \mathbf{f}_j^e$
15:      $\theta_j \leftarrow \theta_j - \gamma \frac{\nabla \mathbf{f}_j}{\|\nabla \mathbf{f}_j\|}$
16:    **end while**
17:    $\theta \leftarrow \theta_j$
18:  **end for**
19: **end for**
20: $G_\theta \leftarrow$ Fit each cluster's set of weight vectors $\theta$ into t-copula distribution

---

*F. Trajectory Generation*

In the previous section, we generated the cost function distribution that best reflects the driver's preferences using the demonstrated driving data. We will then use the learned cost function distribution to derive driver-specific vehicle trajectories with the nonlinear model predictive control (NMPC) strategy because of the nonlinearity imposed by the learned cost function. If the prediction time horizon is small enough, it is assumed that the driver can properly estimate the motion of the preceding vehicle in the longitudinal driving scenario [14]. For each trajectory segment $j$, the optimization problem needs to be solved recursively at each time instance $k$ within the prediction time horizon $T_P$ to generate the driver-specific trajectories, as shown below

$$\begin{aligned} a_0^*(k) &= \arg\min J_j(a_0(k)) \\ J_j &= \theta_j^T \mathbf{f}_j \\ \text{s.t.} \\ d_s &\leq d_0(k), \quad v_{\min} \leq v_0(k) \leq v_{\max} \end{aligned} \quad (16)$$

where $a_0^*(k)$ is the optimal acceleration; $d_s$ is the minimum car-following gap distance that ensures the safety clearance; $v_{\min}$ and $v_{\max}$ are the minimum and maximum longitudinal velocity constraints; $\theta_j$ and $\mathbf{f}_j$ are the feature weight vector sample from the distribution $G_\theta$ and the feature vector based on the observed driving conditions within the planning horizon, respectively, as defined in Sections II-B and II-D. The proposed NMPC design solves (16) with the sequential quadratic programming (SQP) algorithm [37] and only the first value of the predicted acceleration vector $a_{0,k}^P = [a_{0,k}^{P,k}, a_{0,k+\Delta t_s}^{P,k}, \ldots, a_{0,k+T_P-\Delta t_s}^{P,k}]$ is applied to update the vehicle state at the time $k + \Delta t_s$ where $a_0^*(k) = a_{0,k}^{P,k}$.

At the time $k + \Delta t_s$, the vehicle state updating is accomplished using the discretized inter-vehicle dynamics model [38] with sample time $\Delta t_s$ as formulated below

$$\begin{aligned} d_0(k+\Delta t_s) &= (v_{PV}(k) - v_0(k))\Delta t_s + d_0(k) \\ v_0(k+\Delta t_s) &= v_0(k) + a_0^*(k)\Delta t_s \end{aligned} \quad (17)$$

where $v_0$ and $v_{PV}$ are the longitudinal velocity of the human-driven leader truck and preceding vehicle, respectively.

## III. DISTRIBUTED STOCHASTIC MODEL PREDICTIVE CONTROL (DSMPC) DESIGN

In this section, a serial distributed stochastic model predictive control (DSMPC) for the following automated trucks under PF communication technology will be formulated. The following automated truck 1 solves the local optimal control by utilizing the predicted acceleration information of the human-driven leader truck through the learned cost function distribution (16), as motivated from [39]. The remaining following automated trucks solve the local optimal control by utilizing the probability distribution of the predicted acceleration information from its predecessor through V2V communication.

The following objectives are considered for developing the DSMPC strategy for the following automated trucks in the platoon:

1. The local controllers should be recursive feasible and guarantee chance constraint satisfaction for the closed-loop system.
2. The platoon should be $L^\infty$ string stable. This indicates that the peak magnitude of the spacing error should not be amplified through the vehicular string such that $\|\Delta d_{i+1}\|_\infty \leq \|\Delta d_i\|_\infty$, where $\|\Delta d_i\|_\infty$ defines the $L^\infty$ norm of the spacing error of the following truck $i$, as formulated in [22].

A. *Problem Formulation*

In this study, the control objective of the following automated trucks is to regulate the inter-platoon distance gap with a constant distance policy and to maintain zero relative speed to their predecessors in the platoon.

At the time $k$, the vehicle dynamics of each following automated truck in the platoon can be formulated by the nonlinear third-order model [40] as

$$\begin{aligned} \dot{d}_i(k) &= v_{i-1}(k) - v_i(k) \\ \dot{v}_i(k) &= a_i(k) \\ \dot{a}_i(k) &= f_i(v_i, a_i) + g_i(v_i)\varepsilon_i \end{aligned} \quad (18)$$

where $i$ represents the $i$th following automated vehicle in the platoon, $d_i$ defines the inter-vehicle distance, $v_i$ and $a_i$ define longitudinal velocity and acceleration, respectively, $\varepsilon_i$ represents engine input, $f_i(.,.)$ and $g_i(.)$ are formulated as

$$f_i(v_i, a_i) = -\frac{1}{\tau_i}\left(a_i + \frac{\upsilon A_{F_i} C_{d_i}}{2m_i}v_i^2 + \frac{\varphi_{m_i}}{m_i}\right)$$
$$-\frac{\upsilon A_i C_{di} v_i a_i}{m_i} \quad (19)$$
$$g_i(v_i) = \frac{1}{m_i \tau_i}$$

where $\tau_i$ is the actuation time-lag; $\upsilon$ is the mass of the air; $m_i$, $A_{F_i}$, $\varphi_{m_i}$, $C_{d_i}$ are the mass, cross-sectional area, mechanical drag, and drag coefficient of the vehicle $i$, respectively. By applying the control law in [40], the engine input $\varepsilon_i$ can be derived as

$$\varepsilon_i = u_i m_i + \frac{\upsilon A_{F_i} C_{d_i}}{2} v_i^2 + \varphi_{m_i} + \tau_i \upsilon A_{F_i} C_{d_i} v_i a_i \quad (20)$$

where $u_i$ is the control input. After applying the feedback linearization technique [41], $\dot{a}_i(k)$ can be formulated as

$$\dot{a}_i(k) = \frac{1}{\tau_i}\left(u_i(k) - a_i(k)\right) \quad (21)$$

Since the control objective of each following truck in the platoon is to maintain the constant inter-vehicle distance and zero relative speed to its predecessor, the system error dynamics of each following truck can be derived as

$$\begin{aligned} \Delta d_i(k) &= d_i(k) - d_e \\ \Delta v_i(k) &= v_{i-1}(k) - v_i(k) \end{aligned} \quad (22)$$

where $\Delta d_i$ is the deviation from the equilibrium spacing $d_e$ and $\Delta v_i$ is the relative speed of the vehicle $i$ to its predecessor. The system state of each following truck can be defined as $x_i(k) = [\Delta d_i(k), \Delta v_i(k), a_i(k)]^T$ and the state-space form can be defined as

$$\dot{x}_i(k) = A_i x_i(k) + B_i u_i(k) + D_i a_{i-1}(k) \quad (23)$$

where $A_i = \begin{pmatrix} 0 & 1 & 0 \\ 0 & 0 & -1 \\ 0 & 0 & -1/\tau \end{pmatrix}$, $B_i = \begin{pmatrix} 0 \\ 0 \\ 1/\tau \end{pmatrix}$, $D_i = \begin{pmatrix} 0 \\ 1 \\ 0 \end{pmatrix}$.

The discrete version of (23) with a zero-order hold (ZOH) approach can be defined as

$$x_i(k + \Delta t_s) = A_i' x_i(k) + B_i' u_i(k) + D_i' a_{i-1}(k) \quad (24)$$

where $A_i'$, $B_i'$ and $D_i'$ can be obtained by applying Jordan-Chevalley decomposition [42].

The acceleration of the vehicle $i-1$ is considered as an additive disturbance and the state prediction $x_{i,k+\Delta t_s}^{P,k}$ at the time $k + \Delta t_s$ can be derived as

$$x_{i,k+\Delta t_s}^{P,k} = A_i' x_{i,k}^{P,k} + B_i' u_{i,k}^{P,k} + D_i' \tilde{a}_{i-1,k}^{P,k} \quad (25)$$

where the longitudinal acceleration of the preceding truck $\tilde{a}_{i-1,k}^{P,k}$ has a truncated normal distribution $N_{i-1,k}^w$ with sequence $\{\tilde{a}_{i-1,k}^{P,s,k}, \tilde{a}_{i-1,k+\Delta t_s}^{P,s,k}, ..., \tilde{a}_{i-1,k+\Delta t_s}^{P,s,k}, ..., \tilde{a}_{i-1,k}^{P,s+\xi,k}, \tilde{a}_{i-1,k+\Delta t_s}^{P,s+\xi,k}, ..., \tilde{a}_{i-1,k+T_p}^{P,s+\xi,k}\}$ has non-zero mean $\mathrm{E}\left[\tilde{a}_{i-1,k}^{P,k}\right] = \bar{a}_{i-1,k}^{P,k}$ and variance $\Sigma_{i,k}^w$, $\xi$ and $s$ are the number of generated samples and the current generated sample at the time $k$, respectively, and $T_p$ defines the prediction time horizon.

The variables $z_{i,k}^{P,k}$, $e_{i,k}^{P,k}$, $v_{i,k}^{P,k}$ and $K_f$ represent the nominal state, error corresponding to the effect of the disturbances, the nominal control input, and the linear feedback gain, respectively, as shown below

$$\begin{aligned} x_{i,k}^{P,k} &= z_{i,k}^{P,k} + e_{i,k}^{P,k} \\ u_{i,k}^{P,k} &= K_f e_{i,k}^{P,k} + v_{i,k}^{P,k} \end{aligned} \quad (26)$$

The predicted nominal state and the error at the time $k + \Delta t_s$ can be derived as

$$z_{i,k+\Delta t_s}^{P,k} = A_i' z_{i,k}^{P,k} + B_i' v_{i,k}^{P,k}$$
$$e_{i,k+\Delta t_s}^{P,k} = (A' + B'K_f) e_{i,k}^{P,k} + D_i' \tilde{a}_{i-1,k}^{P,k} \quad (27)$$

initialized at $z_{i,k}^{P,k} = x_{i,k}^{P,k}$. The predefined linear feedback gain $K_f$ is used to reduce the growth of the error in the prediction where the optimal control problem is carried out over the nominal control input.

The receding horizon stochastic optimal control problem of each following truck in the platoon at the time $k$ is formulated as in (28)

$$v = \arg\min \left( \mathrm{E}\left[ l_i\left( x_{i,k}^P, v_{i,k}^P \right) \right] \right) \quad (28a)$$

s.t.
$$z_{i,k+\Delta t_s}^{P,k} = A_i' z_{i,k}^{P,k} + B_i' v_{i,k}^{P,k} \quad (28b)$$
$$e_{i,k+\Delta t_s}^{P,k} = (A' + B'K_f) e_{i,k}^{P,k} + D_i' \tilde{a}_{i-1,k}^{P,k} \quad (28c)$$
$$x_{i,k}^{P,k} = z_{i,k}^{P,k} + e_{i,k}^{P,k} \quad (28d)$$
$$u_{i,k}^{P,k} = K_f e_{i,k}^{P,k} + v_{i,k}^{P,k} \quad (28e)$$
$$\tilde{a}_{i-1,k}^{P,k} = \begin{cases} \tilde{a}_{i-1,k}^{P,s,k}, \tilde{a}_{i-1,k+\Delta t_s}^{P,s,k}, \dots, \tilde{a}_{i-1,k+T_P}^{P,s,k}, \dots, \\ \tilde{a}_{i-1,k}^{P,s+\xi,k}, \tilde{a}_{i-1,k+\Delta t_s}^{P,s+\xi,k}, \dots, \tilde{a}_{i-1,k+T_P}^{P,s+\xi,k} \end{cases} \sim \mathrm{N}_{i-1,k}^w \quad (28f)$$
$$\mathrm{P_r}\left[ H_j x_{i,k}^{P,k} \leq h_j \right] \geq 1 - \beta \quad (28g)$$
$$u_{i_{\min}} \leq u_{i,k+n\Delta t_s}^P \leq u_{i_{\max}} \quad (28h)$$
$$z_{i,k+n\Delta t_s}^P \in Z_{i,k} \quad (28i)$$
$$v_{i,k+n\Delta t_s}^P \in \Pi_{i,k} \quad (28j)$$
$$z_{i,k+T_P}^{P,k} \in Z_{i,f} \quad (28k)$$
$$\forall n \in \{0,1,2,\dots,(T_P/\Delta t_s)-1\} \quad (28l)$$
$$\forall s \in \{1,2,\dots,\xi\} \quad (28m)$$
$$\forall k \in \{0,\Delta t_s, 2\Delta t_s,\dots,T\} \quad (28n)$$

where $\mathrm{E}\left[ l_i\left( x_{i,k}^P, v_{i,k}^P \right) \right]$ is the expected value of the cost function; (28g) is the state chance constraints for regulating the vehicle states, such as regulating the maximum deviation from the equilibrium spacing and acceleration constraints; (28h) is used to guarantee the control input values within the predefined range; (28k) is the terminal state constraint to maintain the terminal state to be close enough to the equilibrium point.

**Cost function:** The expected value of the cost function $J_{SMPC_i} = \mathrm{E}\left[ l_i\left( x_{i,k}^P, v_{i,k}^P \right) \right]$ at the time $k$ is derived as

$$J_{SMPC_i} = \mathrm{E}\left[ l_i\left( x_{i,k}^P, v_{i,k}^P \right) \right]$$
$$= \mathrm{E}\left[ \sum_{n=0}^{(T_P/\Delta t_s)-1} \left( \left(x_{i,k+n\Delta t_s}^{P,k}\right)^T Q_x x_{i,k+n\Delta t_s}^{P,k} + \left(u_{i,k+n\Delta t_s}^{P,k}\right)^T R_u u_{i,k+n\Delta t_s}^{P,k} \right) \right.$$
$$\left. + \left(x_{i,k+T_P}^{P,k}\right)^T Q_P x_{i,k+T_P}^{P,k} \right]$$

$$= \sum_{n=0}^{(T_P/\Delta t_s)-1} \left( \begin{matrix} \left(z_{i,k+n\Delta t_s}^{P,k}\right)^T Q_x z_{i,k+n\Delta t_s}^{P,k} + 2\left(z_{i,k+n\Delta t_s}^{P,k}\right)^T Q_x \mathrm{E}\left[e_{i,k+n\Delta t_s}^{P,k}\right] + \\ \left(v_{i,k+n\Delta t_s}^{P,k}\right)^T R_u v_{i,k+n\Delta t_s}^{P,k} + 2\left(v_{i,k+n\Delta t_s}^{P,k}\right)^T R_u K_f \mathrm{E}\left[e_{i,k+n\Delta t_s}^{P,k}\right] \end{matrix} \right) \quad (29)$$
$$+ \left(z_{i,k+T_P}^{P,k}\right)^T Q_P z_{i,k+T_P}^{P,k} + 2\left(z_{i,k+T_P}^{P,k}\right)^T Q_P \mathrm{E}\left[e_{i,k+T_P}^{P,k}\right] + c$$

where $c = \mathrm{E}\left[ \sum_{n=0}^{(T_P/\Delta t_s)-1} \left( \left(e_{i,k+n\Delta t_s}^{P,k}\right)^T (Q_x + K_f^T R_u K_f) e_{i,k+n\Delta t_s}^{P,k} \right) + \left(e_{i,k+T_P}^{P,k}\right)^T Q_P e_{i,k+T_P}^{P,k} \right]$ is a constant term that can be excluded from the cost function since it does not depend on the decision variable $v_{i,k}^{P,k}$; the expected value of the errors $\mathrm{E}\left[e_{i,k+n\Delta t_s}^{P,k}\right]$ and $\mathrm{E}\left[e_{i,k+T_P}^{P,k}\right]$ can be approximated as $D_i' \bar{a}_{i-1,k+n\Delta t}^{P,k}$ and $D_i' \bar{a}_{i-1,k+T_P}^{P,k}$, respectively; $Q_x$ is positive definite weight matrix

$$Q_x = \begin{pmatrix} \psi_1 & 0 & 0 \\ 0 & \psi_2 & 0 \\ 0 & 0 & \psi_3 \end{pmatrix}; R_u \text{ is positive definite weight value and}$$

$Q_P$ is the terminal cost weight which is the solution of the Lyapunov equation $(A' + B'K_f) Q_P (A' + B'K_f) - A' + B'K_f = -(Q_x + K_f R_u K_f^T)$ [43] and the linear feedback gain $K_f$ can be obtained solving the linear quadratic regulator (LQR) optimal control problem.

**State constraints:** The stochastic nature of the longitudinal acceleration of the preceding truck $\tilde{a}_{i-1,k}^{P,k}$ may enforce state constraints violations when defined as deterministic and hard constraints. This subsequently will incur safety-critical issues. Therefore, the state chance constraints (28g) are formulated as in (30), requiring the probability $\mathrm{P_r}(.)$ of violating the constraints based on a risk level $\beta \in (0,1]$

$$\mathrm{P_r}\left[ a_{i,k+s} \geq a_{i_{\min}} \right] \geq 1 - \beta \quad (30a)$$
$$\mathrm{P_r}\left[ a_{i,k+s} \leq a_{i_{\max}} \right] \geq 1 - \beta$$

$$\mathrm{P_r}\left[ \Delta d_{i,k+\sigma,s+\delta} \geq \Delta d_{i,k+\Delta t_s, s+1}^- \right] \geq 1 - \beta \quad (30b)$$
$$\mathrm{P_r}\left[ \Delta d_{i,k+\sigma,s+\delta} \leq \Delta d_{i,k+\Delta t_s, s+1}^+ \right] \geq 1 - \beta$$

$$\Delta d_{i,k+\Delta t_s, s+1}^- = \begin{cases} -d_m & \text{for } i=1 \\ -\max\left( \max\left( \left| \Delta d_{i-1,\sigma}^p \right| \right) \right)_\delta & \text{for } i>1 \end{cases}$$

$$\Delta d_{i,k+\Delta t_s, s+1}^+ = \max\left( \max\left( \left| \Delta d_{i-1,\sigma}^p \right| \right) \right)_\delta \quad \text{for } i>1 \quad (30c)$$

$$\forall \sigma \in \{0, \Delta t_s, 2\Delta t_s, \dots, k+\Delta t_s\}, \forall \delta \in \{1,2,3,\dots,s+1\}$$

where the constraints on the vehicle acceleration are imposed for the vehicle drivability in (30a); $d_m$ is a predefined spacing error. The constraint of the spacing error guarantees the string stability $L^\infty$ with the probability as formulated in (30b) and (30c), which will be proved later in the following section. The proposed DSMPC design uses the implementation as introduced in Algorithm 2.

| Algorithm 2: The DSMPC algorithm |
|---|
| **Input:** The human-driven leader truck's predicted distribution of the longitudinal acceleration |
| **Output:** The following trucks' trajectories |
| 1: At time $k = 0$, |
| 2:   Initialize $x_{i,0} = 0$ for all the following trucks |
| 3: At time $k > 0$ |
| 4:   **for** $i = 1$ |
| 5:     The following truck $i$ predicts the set of the longitudinal acceleration with $\xi$ samples of the human-driven leader truck $S_{a_{0,k}^P}$ by solving (16) and fits $S_{a_{0,k}^P}$ to the non-zero mean normal distribution $N_{0,k}^w$ bounded with $\left[\min\left(S_{a_{0,k}^P}\right), \max\left(S_{a_{0,k}^P}\right)\right]$ |
| 6:     The following truck $i$ derives the $\xi$ samples of the predicted vehicle state $S_{x_{i,k}^P}$ by solving (28) |
| 7:     Following truck $i$ transmits $S_{a_{i,k}^P}$ to $i+1$ |
| 8:   **end** |
| 9:   **for** $1 < i \leq 3$ |
| 10:    The following truck $i$ receives the $\xi$ samples of predicted longitudinal acceleration values $S_{a_{i-1,k}^P}$ from the following truck $i-1$ and fits $S_{a_{i-1,k}^P}$ to the non-zero mean normal distribution $N_{i-1,k}^w$ bounded with $\left[\min\left(S_{a_{i-1,k}^P}\right), \max\left(S_{a_{i-1,k}^P}\right)\right]$ |
| 11:    The following truck $i$ derives the $\xi$ samples of the predicted vehicle state $S_{x_{i,k}^P}$ by solving (28) |
| 12:    **if** $i < 3$ |
| 13:      Following truck $i$ transmits $S_{a_{i,k}^P}$ to $i+1$ |
| 14:    **end** |
| 15:   **end** |
| 16: Update time $k = k + \Delta t_s$ and go to step 3 |

**Terminal constraint:** The terminal constraint set $Z_{i,f}$ should be considered as the subset of the robust positive invariant set $Z_{i,f} \subseteq Z_{RPI}$ to guarantee that each local controller will remain in the robust invariant set once it enters, as discussed in [22]. By this, a terminal control law $\pi_f\left(z_{i,k+T_P}^{P,k}\right) \in \Pi_{i,\infty}$ is required when each local controller reaches the robust positive invariant set $Z_{RPI}$. The terminal control law can be constructed with the obtained linear feedback gain such as $\pi_f\left(z_{i,k+T_P}^{P,k}\right) = K_f z_{i,k+T_P}^{P,k}$ and we have $A_i' z_{i,k+T_P}^{P,k} + B_i' \pi_f\left(z_{i,k+T_P}^{P,k}\right) \in Z_{i,f}$ for all $z_{i,k+T_P}^{P,k} \in Z_{i,f}$ under the terminal control law.

*B. Platoon Feasibility, Chance Constraint Satisfaction and String Stability Analysis*

In the previous section, we formulated the DSMPC as a formation control strategy for the following automated trucks in the human-leading heavy-duty truck platoon. Next, we will establish the recursive feasibility, chance constraint satisfaction, and string stability of the proposed DSMPC design with the following lemmas and proofs.

*Lemma 1:* The DSMPC algorithm for the following trucks is recursive feasible, i.e. it is feasible for all times $0 \leq k \leq T - T_p$ if the initial state is feasible and the following states remain in the feasible set.

**Proof:** Since the stochastic variables in the optimization problem of the proposed DSMPC design only affect the cost, the recursive feasibility can be established in terms of the deterministic nominal state and control input using the traditional principles in predictive control, as stated in [44]. Assume that DSMPC is initially feasible with respect to the constraints (28, 30) at $k = 0$, and there exists an optimal sequence for all following trucks $v_{i,0}^P = \left[v_{i,0}^{P,0}, v_{i,\Delta t_s}^{P,0}, ..., v_{i,T_P-\Delta t_s}^{P,0}\right] \in \Pi_{i,0}$ with the resulting nominal state trajectory $z_{i,0}^P = \left[z_{i,\Delta t_s}^{P,0}, z_{i,2\Delta t_s}^{P,0}, ..., z_{i,T_P}^{P,0} \in Z_{i,0}\right]$ and $z_{i,T_P}^{P,0} \in Z_{i,f}$ where $Z_{i,f}$ is the terminal state domain. Next time interval at $k = \Delta t_s$, the solution is also feasible for the following trucks with the optimal sequence $v_{i,\Delta t_s}^P = \left[v_{i,\Delta t_s}^{P,1}, v_{i,2\Delta t_s}^{P,1}, ..., v_{i,T_P}^{P,1}, K_f z_{i,T_P}^{P,1}\right] \in \Pi_{i,\Delta t_s}$; $z_{i,\Delta t_s}^P = \left[z_{i,2\Delta t_s}^{P,1}, z_{i,3\Delta t_s}^{P,1}, ..., z_{i,T_P}^{P,1}, A_i' z_{i,T_P}^{P,1} + B_i' K_f z_{i,T_P}^{P,1}\right] \in Z_{i,\Delta t_s}$ and $A_i' z_{i,T_P}^{P,1} + B_i' K_f z_{i,T_P}^{P,1} \in Z_{i,f}$. Therefore, the DSMPC algorithm is feasible at all time steps using induction.

*Lemma 2:* The DSMPC algorithm for the following trucks satisfy the closed-loop chance constraints (28g) for $k = 0,...,T$ and $j = 1,...,\varsigma$ if and only if the nominal system (28b) satisfies the constraints (28i) with

$$Z_{i,k} = \left\{z \in \mathbb{R}^n \mid H_j z_{i,k}^{P,k} \leq \eta_{j,i,k}\right\} \tag{31}$$

where $\eta_{j,i,k}$ is given by

$$\begin{aligned}\eta_{j,i,k} &= \max_{\eta} \eta \\ \text{s.t.} \quad & \\ P_r\left[\eta_{j,i,k} \leq h_j - H_j e_{i,k}^{P,k} - E\left(D_i' \tilde{a}_{i-1,k}^{P,k}\right)\right] &\geq 1 - \beta\end{aligned} \tag{32}$$

**Proof:** The state chance constraints (28g) can be expressed with the predicted nominal state and the stochastic disturbance as shown in (33)

$$P_r\left[H_j z_{i,k}^{P,k} \leq h_j - H_j e_{i,k}^{P,k}\right] \geq 1 - \beta$$

or
$$\text{P}_r\left[\eta_{j,i,k} \le h_j - H_j e^{P,k}_{i,k}\right] \ge 1-\beta \quad (33)$$

where $\eta_{j,i,k}$ denotes a bound that is calculated as $H_j z^{P,k}_{i,k} \le \eta_{j,i,k}$ and (33) can be reformulated as shown in (34)

$$\text{P}_r\left[\eta_{j,i,k} \le h_j - H_j e^{P,k}_{i,k} - \text{E}\left(D_i' \tilde{a}^{P,k}_{i-1,k}\right)\right] \ge 1-\beta \quad (34)$$

where the scalar value $\text{E}\left(D_i' \tilde{a}^{P,k}_{i-1,k}\right)$ defines an additional constraint tightening to deal with the non-zero mean $\bar{a}^{P,k}_{i-1,k}$. The chance constraint in (34) can be expressed in terms of the cumulative density function (CDF) $F_H$ of the random variable $h_j - H_j e^{P,k}_{i,k} - \text{E}\left(D_i' \tilde{a}^{P,k}_{i-1,k}\right)$ as shown in (35)

$$F_H\left(-\eta_{j,i,k}\right) \ge 1-\beta \quad (35)$$

where the probability of the truncated distribution $\text{N}^w_{i-1,k}$ is known. Thus, the value of $\eta_{j,i,k}$ can be calculated by evaluating the inverse CDF $F_H^{-1}$ as shown in (36)

$$-\eta_{j,i,k} = -F_H^{-1}\left(1-\beta\right) \quad (36)$$

which concludes the proof.

***Lemma 3:*** The DSMPC algorithm for the following trucks is $L^\infty$ string stable when the DSMPC algorithm is recursive feasible, and the chance constraints are satisfied.

**Proof:** When the DSMPC algorithm is recursive feasible and the chance constraints are satisfied, the state constraint on the spacing error (30b) is satisfied at all time steps. Therefore, the $L^\infty$ string stability $\|\Delta d_{i+1}\|_\infty \le \|\Delta d_i\|_\infty$ is satisfied when $k \to \infty$ [22].

## IV. RESULTS AND DISCUSSION

In this section, the performance of the proposed stochastic driver behavior model and the truck platooning control strategy will be investigated. Besides, a comparison study with a deterministic formation-controlled platoon will be conducted to evaluate the controller performance of the proposed design.

### A. Stochastic Driver Behavior Model Implementation

In this study, we used a driver-in-the-loop driving simulation integrated with MATLAB Automated Driving Toolbox to collect realistic driving data for training and testing the proposed stochastic driver behavior model. The setup consists of a driving seat, three curved monitors, a steering wheel, and pedals is provided, as shown in Fig. 3. The simulation cases focus on the single-lane longitudinal driving scenarios where a heavy-duty truck driver follows a preceding vehicle that is operated under several predefined speed profiles. Fig. 4 shows an example scene from the 3D simulation environment where a driver (blue truck) follows a preceding vehicle (red vehicle) on the road. The driving data set is collected under 10 different driving scenarios, and each driving scenario is performed 30 different times by three drivers. The driver model is developed using 300 leader-follower trajectories for each driver. The data is obtained from the simulation environment at 10 Hz.

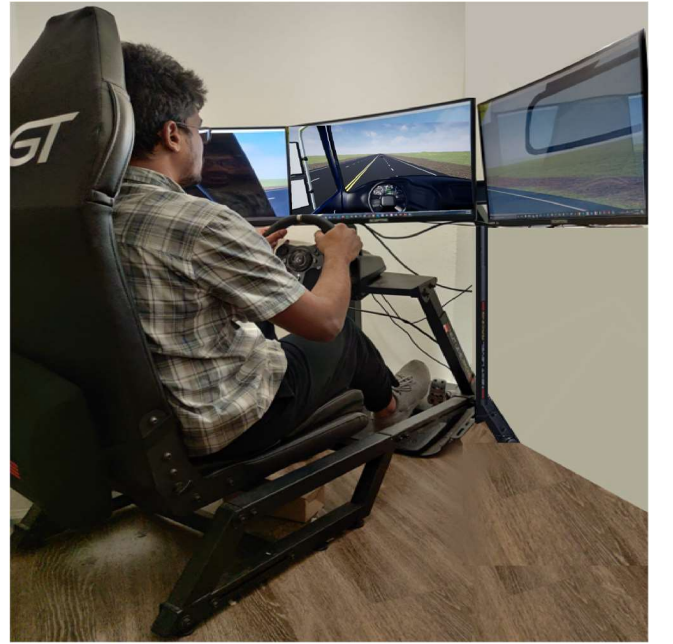

Fig. 3. A driving scene when a driver operates the truck in the 3D simulation environment.

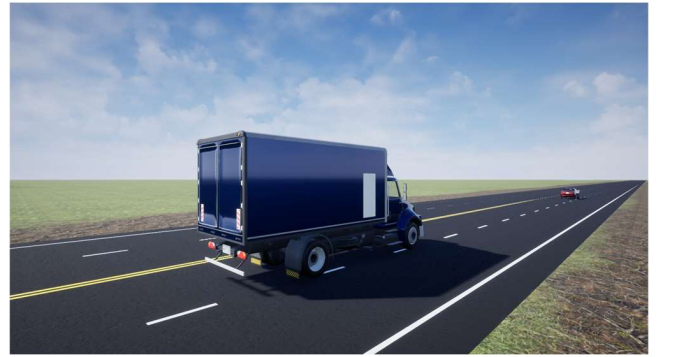

Fig. 4. A road scene from the 3D simulation environment.

To assess the performance of the proposed stochastic driver behavior model for each driver, 25 trajectories for each driving situation are chosen randomly for training, and the remaining 5 trajectories for each driving scenario are used for testing. In the weight vector update, the learning rate $\gamma$ is set to 0.2 at the initial and then drops by half for every five epochs. The length of each trajectory segment $T_H$ and subsegment $T_P$ are set to 3 seconds and 1 second, respectively. The sample time $\Delta t_s$ is 0.1 seconds, and the safe car-following gap distance between the vehicles $d_s$ is set to 5 m. In the trajectory optimization with respect to quintic spline coefficients, as in step 11 of Algorithm 1, the BFGS Quasi-Newton method is used [45]. Besides, $v_{\min}$ is set to 0 m/s and

$v_{max}$ is set to the maximum speed of the preceding traffic during the trip in (16). For the trajectory generation, 50 samples for each driving scenario are generated by using the NMPC design, as defined in (16).

### B. Stochastic Driver Behavior Model Assessment

In this section, we will evaluate the performance of the proposed stochastic driver behavior model in various driving scenarios. Fig. 5 shows the $L^2$ norm of weight update gradients for the trajectory segments used in training for Driver 1. Notably, the feature weights converge after roughly 40 epochs for all the trajectory segments during the optimization. For demonstration purposes, we present the graphical results of one trajectory (Fig. 6, Fig. 7, and Fig. 8) used for testing for Driver 1 and summarize the numerical results from all other trajectories for each driver in Table I. The predicted one sample and actual one sample of a driving scenario in testing are shown in Fig. 6 and Fig. 7. The predicted 50 samples and the actual 5 samples of this driving scenario in testing are plotted in Fig. 8. The results show that the stochastic driver behavior model can generate diverse trajectory samples which represent the richness of the driver's unique driving preferences. Furthermore, it can be seen that the proposed stochastic driver behavior model generates accurate trajectories, although the model has never seen the testing trajectories previously, suggesting a high degree of validity.

Root Mean Square Error (RMSE) between predicted and actual trajectories is used to evaluate the performance of the learning model. Table I shows the average RMSE values between the mean of the actual and predicted trajectories among all driving scenarios for each driver. The results show that the proposed driver behavior model can mimic the observed trajectories with small prediction errors for each driver. Because the expected features are determined as the feature values of the most likely trajectories throughout the learning process, small deviations between the actual and predicted trajectories are acceptable in testing.

To conclude the discussion in the driver behavior model evaluation, the findings show that the proposed stochastic driver behavior model is transferable to different truck drivers and it can learn and imitate each driver's observed unique and stochastic driving behaviors in various driving scenarios.

TABLE I
AVERAGE RMSE VALUES IN TESTING

| Truck Driver | Speed (m/s) | Acceleration (m/s²) |
|---|---|---|
| Driver 1 | 1.10 | 0.50 |
| Driver 2 | 1.40 | 0.73 |
| Driver 3 | 1.28 | 0.54 |

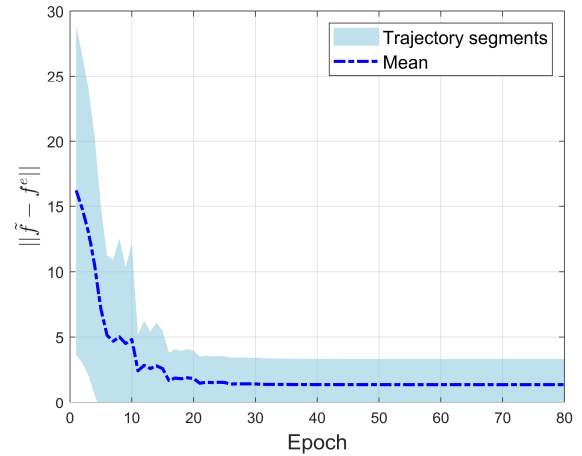

Fig. 5. Gradients ($L^2$ norm) of the trajectory segments for training.

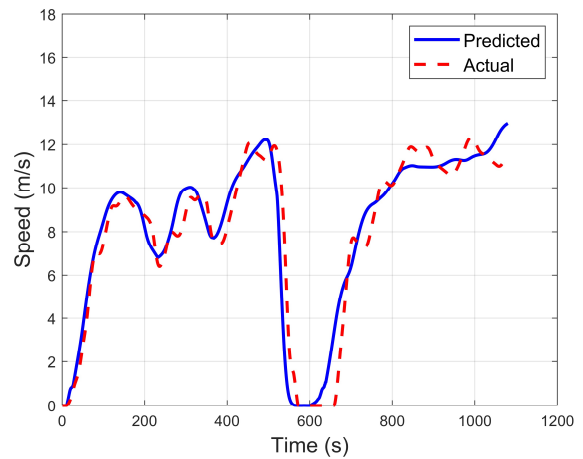

Fig. 6. Speed trajectories for testing in a driving scenario (one sample).

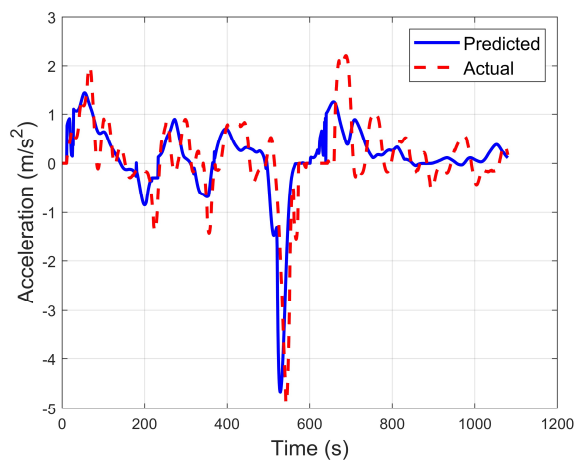

Fig. 7. Acceleration trajectories for testing in a driving scenario (one sample).

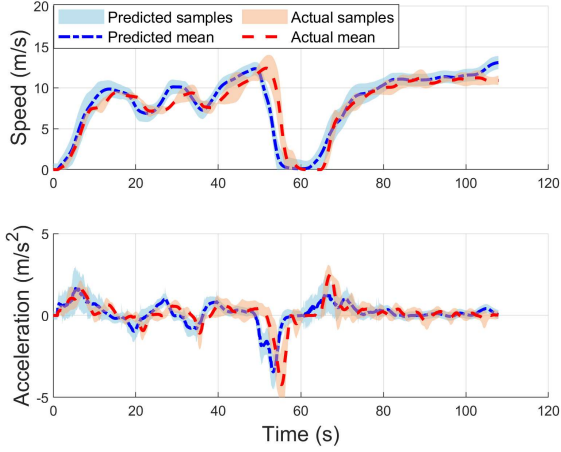

Fig. 8. Speed and acceleration trajectories for testing in a driving scenario (all samples).

## C. Baseline Deterministic Platoon Control Design

We compare our proposed DSMPC design with a serial distributed deterministic model predictive control design (DMPC) to analyze the controller performance of the truck platoon. The DMPC design is a well-established and widely accepted distributed formation control approach to control the vehicular platoon in a form [20-23]. The baseline DMPC formation control strategy is adopted for each following automated truck to maintain the constant gap distance and zero speed difference to their predecessors in the platoon. In the deterministic DMPC baseline strategy, the uncertainty of the longitudinal acceleration of the human-driven leader truck is ignored and only the expected mean value $\mathrm{E}\left[\tilde{a}_{i-1,k}^{P,k}\right] = \bar{a}_{i-1,k}^{P,k}$ is considered. The baseline DMPC design uses the implementation strategy as introduced in Algorithm 3.

In the baseline DMPC design, the receding horizon deterministic optimal control problem of each following truck in the platoon at the time $k$ is formulated as:

$$u = \arg\min J_{MPC_i}\left(x_{i,k}^{P}, u_{i,k}^{P}\right)$$

$$J_{MPC_i} = \sum_{n=1}^{T_P/\Delta t_s}\left[\left(x_{i,k+n\Delta t_s}^{P,k}\right)^T Q_x x_{i,k+n\Delta t_s}^{P,k} + R_u\left(u_{i,k+(n-1)\Delta t_s}^{P,k}\right)^2\right]$$
$$+\left(x_{i,k+T_P}^{P,k}\right)^T Q_P x_{i,k+T_P}^{P,k} \quad (37a)$$

s.t.

$$x_{i,k+\Delta t_s}^{P,k} = A_i' x_{i,k}^{P,k} + B_i' u_{i,k}^{P,k} + D_i' \bar{a}_{i-1,k}^{P,k} \quad (37b)$$

$$u_{i_{\min}} \leq u_{i,k+n\Delta t_s} \leq u_{i_{\max}} \quad (37c)$$

$$a_{i_{\min}} \leq a_{i,k+n\Delta t_s} \leq a_{i_{\max}} \quad (37d)$$

$$\Delta d_{i,k+\Delta t_s}^{-} \leq \Delta d_{i,k+n\Delta t_s} \leq \Delta d_{i,k+\Delta t_s}^{+} \quad (37e)$$

$$\Delta d_{i,k+\Delta t_s}^{-} \leq \Delta d_{i,k+n\Delta t_s} \leq \Delta d_{i,k+\Delta t_s}^{+}$$

$$\Delta d_{i,k+\Delta t_s}^{-} = \begin{cases} -d_m & \text{for } i=1 \\ -\max\left(\left|\Delta d_{i-1,\sigma}\right|\right) & \text{for } i>1 \end{cases} \quad (37f)$$

$$\Delta d_{i,k+\Delta t_s}^{+} = \max\left(\left|\Delta d_{i-1,\sigma}\right|\right) \quad \text{for } i>1 \quad (37g)$$

$$\forall \sigma \in \{0, \Delta t_s, 2\Delta t_s, ..., k+\Delta t_s\}$$

$$x_{i,k+n\Delta t_s}^{P} \in X_{i,k} \quad (37h)$$

$$u_{i,k+n\Delta t_s}^{P} \in U_{i,k} \quad (37i)$$

$$x_{i,k+T_P}^{P,k} \in X_{i,f} \quad (37j)$$

---

Algorithm 3: The DMPC algorithm
**Input:** The human-driven leader truck's predicted distribution of the longitudinal acceleration
**Output:** The following trucks' trajectories
1: At time $k=0$,
2:     Initialize $x_{i,0} = 0$ for all the following trucks
3: At time $k>0$
4:     **for** $i=1$
5:         The following truck $i$ predicts the set of the longitudinal acceleration with $\xi$ samples of the human-driven leader truck $S_{a_{0,k}^P}$ by solving (16) and fits $S_{a_{0,k}^P}$ to the non-zero mean normal distribution $N_{0,k}^w$ bounded with $\left[\min\left(S_{a_{0,k}^P}\right), \max\left(S_{a_{0,k}^P}\right)\right]$
6:         The following truck $i$ uses the expected mean value of the human-driven leader truck $\mathrm{E}\left[\tilde{a}_{i-1,k}^{P,k}\right] = \bar{a}_{i-1,k}^{P,k}$ derives the predicted vehicle state $x_{i,k}^P$ by solving (37)
7:         Following truck $i$ transmits $a_{i,k}^P$ to $i+1$
8:     **end**
9:     **for** $1 < i \leq 3$
10:        The following truck $i$ receives the predicted longitudinal acceleration $a_{i-1,k}^P$ from the following truck $i-1$ and the following truck $i$ derives the predicted vehicle state $x_{i,k}^P$ by solving (37)
11:       **if** $i<3$
12:           Following truck $i$ transmits $a_{i,k}^P$ to $i+1$
13:       **end**
14:     **end**
15:     Update time $k = k + \Delta t_s$ and go to step 3

---

## D. Evaluation of the DSMPC Design with Comparison Study

One of the driving scenarios from the data set is used to simulate the traffic in front of the truck platoon. The speed profile of this driving scenario can be found in Fig. 9 (preceding traffic). The truck platoon with the leader truck (Driver 1) follows the preceding traffic according to the control objectives as previously introduced in (16) and (28). We carry out 100 Monte Carlo simulations to evaluate the performance of the proposed design. The model parameters of the proposed DSMPC and the baseline DMPC designs are listed in Table II. For the sake of conciseness, the following trucks are abbreviated as "FT1", "FT2", and "FT3", respectively, in the results.

Firstly, we evaluate the performance of the proposed DSMPC design through the number of constraint violations and the spacing error. Table III shows the average constraint violations and maximum spacing error of the proposed DSMPC and DMPC baseline designs for each following truck among the generated 100 simulations. The boxplots of the maximum spacing errors of the following trucks in the platoon are shown in Fig. 10. According to the results in Table III, the proposed DSMPC design achieves the closed-loop constraint satisfaction of 100%, 99.95%, and 99.94% for the following trucks in the platoon. This indicates that the constraints of each following truck are satisfied with the specified probability of 95% in the closed-loop. By incorporating the benefits of the stochastic prediction model and the chance constraint implementation, the proposed DSMPC design can achieve better controller performance with fewer constraint violations and better spacing error tracking compared to the DMPC baseline strategy.

TABLE III
CONSTRAINT VIOLATIONS AND SPACING ERROR COMPARISON

| Truck | Average constraint violations (DSMPC) | Average maximum spacing error (DSMPC) | Average constraint violations (DMPC) | Average maximum spacing error (DMPC) |
|---|---|---|---|---|
| FT1 | 0 | 0.19 m | 0 | 0.78 m |
| FT2 | 0.34 | 0.15 m | 5.80 | 0.28 m |
| FT3 | 0.43 | 0.14 m | 3.73 | 0.28 m |

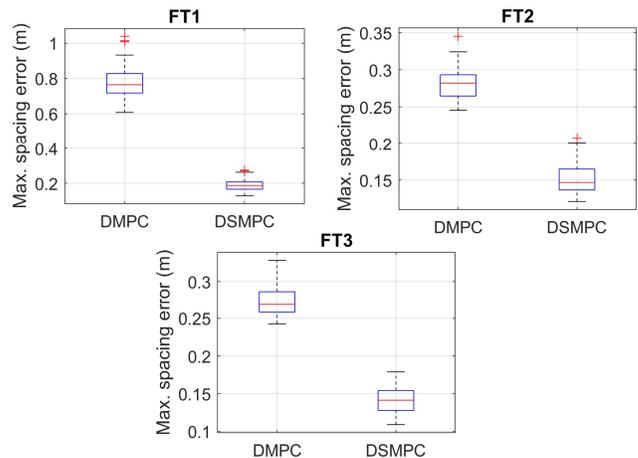

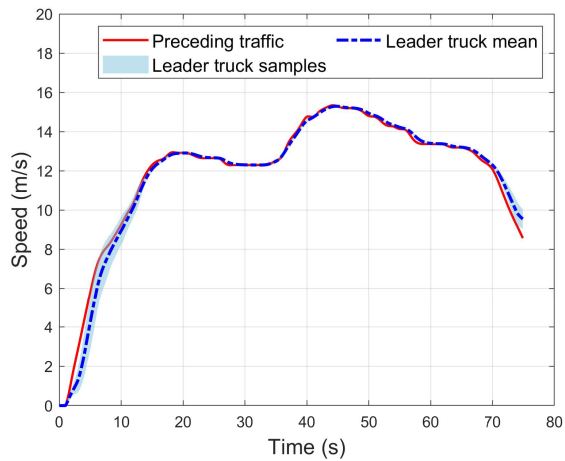

Fig. 9. Speed trajectories of the human-driven leader truck and preceding traffic.

Fig. 10. Boxplots of the maximum spacing errors of the following trucks in the comparison study.

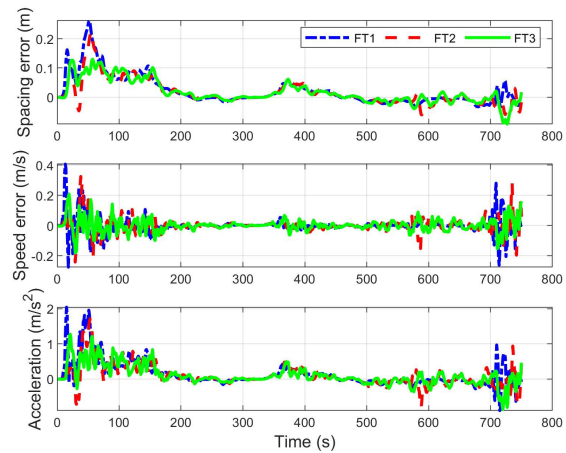

Fig. 11. The speed and spacing errors and acceleration of the following trucks in the platoon with DSMPC design.

TABLE II
DSMPC AND DMPC PARAMETERS

| Parameter | Value | Parameter | Value |
|---|---|---|---|
| $\Delta t_s$ | 0.1 s | $\beta$ | 0.05 |
| $T_P$ | 1 s | $\psi_{1,2,3}$ | 1 |
| $u_{i_{min}}$ | -4 m/s² | $Z_{i,f}, X_{i,f}$ | [0,0,0] |
| $u_{i_{max}}$ | 4 m/s² | $Z_{i,0}, X_{i,0}$ | [0,0,0] |
| $a_{i_{min}}$ | -3 m/s² | $d_m, d_e$ | 3 m, 5 m |
| $a_{i_{max}}$ | 3 m/s² | $\xi, T$ | 10, 75 |
| $R_u$ | 0.5 | $\tau_i$ | 0.45 |

The string stability of the truck platoon is another critical factor in the performance analysis of the proposed control design. Fig. 11 shows the evolution of speed and spacing errors and acceleration of the following trucks in one simulation run. Table IV shows the $L^\infty$ norm of each following truck during the trip in one simulation run. Table V shows the minimum, average, and maximum observed $L^\infty$ norm of each following truck during the trip among generated samples. These findings reveal that the truck platoon is satisfying the $L^\infty$ string stability condition $\|\Delta d_{i+1}\|_\infty \leq \|\Delta d_i\|_\infty$

page13.mdby successfully utilizing the $L^\infty$ norm constraints in (30), and the errors decay throughout the platoon.

TABLE IV
$L^\infty$ Spacing Error Comparison in One Simulation Run

| Truck | $L^\infty$ |
|---|---|
| FT1 | 0.26 m |
| FT2 | 0.22 m |
| FT3 | 0.13 m |

TABLE V
$L^\infty$ Spacing Error Comparison in All Generated Samples

| Truck | $L^\infty_{min}$ | $L^\infty_{avg}$ | $L^\infty_{max}$ |
|---|---|---|---|
| FT1 | 0.13 m | 0.19 m | 0.28 m |
| FT2 | 0.12 m | 0.15 m | 0.21 m |
| FT3 | 0.11 m | 0.14 m | 0.18 m |

To summarize the discussion in the performance evaluation of the proposed DSMPC design, the results show that the proposed DSMPC design can achieve better controller performance compared to the deterministic DMPC baseline method. Furthermore, the proposed DSMPC design can achieve a string stable platoon by satisfying sufficient conditions. The findings demonstrate that the proposed DSMPC design has the advantages of accurately representing stochastic driver behaviors and explicitly treating human-induced uncertainty to navigate the human-leading truck platoon safely.

## V. Conclusion

In this study, we designed a distributed stochastic model predictive control design for a human-leading heavy-duty truck platoon. The proposed control framework integrates a stochastic driver behavior learning model of the human-driven leader truck with a distributed control strategy for the following automated trucks in the platoon. The stochastic driving behaviors of the leader truck are learned by acquiring the cost function distribution from the driver-specific demonstrations with inverse reinforcement learning. In the distributed control design, the following automated trucks predict the distribution of the predecessors' acceleration maneuvers. The proposed stochastic formation control strategy handles the uncertainty of the predicted acceleration information to control the platoon in a form by satisfying the recursive feasibility, chance constraints, and string stability. The comparison results reveal that the human-leading heavy-duty truck platoon with the proposed control strategy can achieve minor constraint violations and spacing errors than the deterministic baseline model predictive control. The results collectively demonstrate that the proposed stochastic distributed control strategy is effective for a human-leading heavy-duty truck platoon, a new form of platoon leveraging vehicle autonomy and human intelligence.

As a further extension of the work, we will investigate different drivers' behaviors of the human-driven leader truck and their impacts on the following automated trucks in the platoon.

Acknowledgment
## Acknowledgment

The authors would like to thank Mr. Abishek Joseph Rocque for his assistance in the driver-in-the-loop simulation design and data collection.

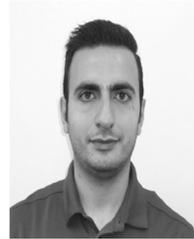

**MEHMET FATIH OZKAN** received the B.E. degree in mechatronics engineering from Mevlana University, Konya, Turkey, in 2016 and the M. S. degree in computer science from Texas A&M University-Corpus Christi, Corpus Christi, TX, USA, in 2019.

He is currently pursuing a Ph.D. degree in mechanical engineering from Texas Tech University, Lubbock, TX, USA. His research interests include connected and automated vehicles, intelligent transportation systems, driver behavior modeling, driver behaviors study, and vehicle system control.

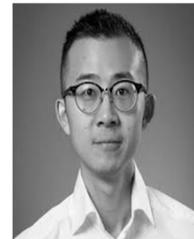

**YAO MA** received the B.E. degree in control science and engineering from the Harbin Institute of Technology, Harbin, China, in 2012, the M.S. degree in electrical and computer engineering from North Carolina State University, Raleigh, NC, USA, in 2013, and the Ph.D. degree in mechanical engineering from the University of Texas, Austin, TX, USA, in 2019.

He is currently an Assistant Professor with the Department of Mechanical Engineering, Texas Tech University, TX, USA. His research interests include connected and automated vehicles, intelligent transportation systems, driver behaviors study, and vehicle system control.